\documentclass[conference,a4paper]{APSIPA2021}
\usepackage{multirow}
\usepackage{graphicx}
\usepackage{amsmath}
\usepackage[psamsfonts]{amssymb}
\usepackage{amsxtra}
\usepackage{threeparttable}
\usepackage{booktabs}

\begin{document}

\title{A Study on Decoupled Probabilistic Linear Discriminant Analysis}

\author{%
\authorblockN{%
Di Wang\authorrefmark{1}\authorrefmark{2}, 
Lantian Li\authorrefmark{1}, 
Hongzhi Yu\authorrefmark{2}, 
Dong Wang\authorrefmark{1}
}
\authorblockA{%
\authorrefmark{1}
Center for Speech and Language Technologies, BNRist, Tsinghua University, China \\
E-mail: \{wangdi,lilt\}@cslt.org; wangdong99@mails.tsinghua.edu.cn}
\authorblockA{%
\authorrefmark{2}
Key Laboratory of China’s Ethnic Languages and Information Technology of Ministry of Education, \\ Northwest Minzu University, China \\
E-mail: yhz@xbmu.cn}
}

\maketitle
\thispagestyle{empty}

\begin{abstract}
Probabilistic linear discriminant analysis (PLDA) has broad application in open-set verification tasks, such as speaker verification.
A key concern for PLDA is that the model is too simple (linear Gaussian) to deal with complicated data;
however, the simplicity by itself is a major advantage of PLDA, as it leads to desirable generalization.
An interesting research therefore is how to improve modeling capacity of PLDA while retaining the simplicity.
This paper presents a decoupling approach, which involves a global model that is simple and generalizable,
and a local model that is complex and expressive. While the global model holds a bird view on the entire data,
the local model represents the details of individual classes.
We conduct a preliminary study towards this direction and investigate a simple decoupling model 
including both the global and local models. 
The new model, which we call decoupled PLDA, is tested on a speaker verification task.
Experimental results show that it consistently outperforms the vanilla PLDA when the model is based on raw speaker vectors.
However, when the speaker vectors are processed by length normalization, the advantage of decoupled PLDA will be largely lost,
suggesting future research on non-linear local models. 

\end{abstract}


\section{Introduction}

Probabilistic linear discriminant analysis (PLDA)~\cite{ioffe2006probabilistic,prince2007probabilistic}
has been widely used as the scoring model in open-set verification tasks, such as speaker verification.
It represents the data with a linear Gaussian model, where the between-class distribution is a Gaussian and
the within-class distributions of individual classes are homogeneous Gaussians, i.e., Gaussians with
a shared covariance.


An obvious weakness of the PLDA model is the underlying linear Gaussian assumption, which is too simple to
represent data with complex distributions. However, simply employing a more complex model
is not always recommended, as the generalization capability to unseen classes may be degraded~\cite{villalba2017tied,li2020neural}.
In fact, simplicity by itself is a key advantage of PLDA. An interesting research topic therefore is
how to improve the model capacity, while retaining its simplicity.

In this paper, we present a decoupling approach, which views the data from two perspectives:
a high-level and global perspective that focuses on the distribution of the entire data, including those unseen classes;
and a low-level and local perspective that scrutinizes the distribution of individual classes.
The high-level perspective relies on a rough model $p_g$, for which generalization to unseen classes is the most important,
while the low-level perspective relies on a detailed model $p_l$, by which an accurate within-class probability can be obtained.
We combine $p_g$ and $p_l$ using the normalized likelihood (NL) formulation of PLDA~\cite{borgstrom2013discriminatively,mccree2017extended,wang2020remarks},
where $p_g$ is employed to perform enrollment and normalization, while $p_l$ is employed to perform prediction.
This leads to a decoupled model that represents the rough/global distribution
and detailed/local in a single statistical framework.

A key advantage of the decoupling approach is that the local model $p_l$ can be very complex and so
can represent complicated data.
Investigation on this capability will be left for future work. In this study, we focus on
a simple case where $p_l$ remains a Gaussian, but independent of the global model $p_g$.
The resultant model will be called \textbf{decoupled PLDA}. We tested the model on a
speaker verification task, and the results demonstrated that even with this simple case,
the decoupling approach can offer interesting performance improvement if the model was based on 
raw speaker vectors, though this advantage was largely lost if the speaker vectors were length normalized~\cite{garcia2011analysis}.


\section{Theory}

\subsection{Revisit PLDA}

PLDA represents data by a linear Gaussian model.
To simplify the presentation, we shall assume that the covariance matrix of the between-class distribution is diagonal
and the shared covariance matrix of the within-class distributions of individual classes is an identify matrix.
This can easily obtained by simultaneous diagonalization with a linear transform $\mathbf{W}$. Put it more formally:

\begin{equation}
p(\pmb{\mu}) = \mathcal{N} (\pmb{\mu}; \pmb{0}, \textnormal{diag}(\pmb{\epsilon}) )
\end{equation}

\begin{equation}
p(\pmb{x}|\pmb{\mu}) = \mathcal{N} (\pmb{x}; \pmb{\mu}, \mathbf{I}).
\end{equation}
\noindent where $\pmb{x}$ has been transformed by $\mathbf{W}$ and $\textnormal{diag}(\pmb{\epsilon})$ is a diagonal matrix constructed from
$\pmb{\epsilon}$.

According to the hypothesis test theory~\cite{neyman1933ix},
the following likelihood ratio (LR) has the highest power among other scores when determining if a test sample $\pmb{x}$
belongs to the class represented by the enrollment samples $\{\pmb{x}_1, ..., \pmb{x}_n\}$:

\begin{equation}
\label{eq:lr}
\emph{LR} = \frac{p(\pmb{x}, \pmb{x}_1, ..., \pmb{x}_n)}{ p(\pmb{x}) p(\pmb{x}_1, ..., \pmb{x}_n)}.
\end{equation}

It has been found that the decision based on PLDA LR is optimal in term of minimum Bayes risk (MBR)~\cite{wang2020remarks}.

The parameters of the model involve $\pmb{\epsilon}$ and $\mathbf{W}$. They are usually optimized by maximizing likelihood (ML) with
the following objective function:

\begin{equation}
\mathcal{L}(\pmb{\epsilon}, \mathbf{W}) = \sum_{k=1}^K \ln p(\pmb{x}^k_1, ..., \pmb{x}^k_{n_k}),
\end{equation}
\noindent where $\pmb{x}^k_{i}$ denotes the $i$-th sample of the $k$-th class.

\subsection{Decoupling form}

The LR form in (\ref{eq:lr}) is equivalent to the following form~\cite{borgstrom2013discriminatively,mccree2017extended,wang2020remarks}:

\begin{equation}
 \emph{NL}=\frac{\int p(\pmb{x}|\pmb{\mu}) p(\pmb{\mu}|\pmb{x}_1, ..., \pmb{x}_n) \rm{d} \pmb{\mu}}{ p(\pmb{x}) }.
\end{equation}

This was denoted by normalized likelihood (NL) in~\cite{wang2020remarks} in order to emphasize on the normalization role of the denominator. 
By this NL form, the LR score is decomposed into three components:

\begin{itemize}
\item The \textbf{enrollment} component $p(\pmb{\mu}|\pmb{x}_1, ..., \pmb{x}_{n})$ that produces the posterior of the class mean $\pmb{\mu}$ given the enrollment data $\pmb{x}_1, ..., \pmb{x}_{n}$.
\item The \textbf{prediction} component $p(\pmb{x}|\pmb{\mu})$ that computes the probability that a test sample $\pmb{x}$ belongs to a class represented by $\pmb{\mu}$.
\item The \textbf{normalization} component $p(\pmb{x})$ that computes the probability that $\pmb{x}$ is produced by all potential classes.
\end{itemize}

Among the three components, the enrollment and normalization concern all potential classes and so are relevant to the global distribution,
while the prediction concerns the distribution of a particular class and so is relevant to the local distribution.
In the vanilla PLDA, all the three components are derived from the same underlying generative process,
and so are \emph{coupled} together.
\textbf{In cases where the generative model matches the data, the coupled model is optimal for scoring in terms of Bayes risk.}

In practice, however, the data could be more complex than what the presumed generative model can represent.
For example, in speaker verification, it was well known that
both the between-class and within-class distributions of speaker vectors
tend to be super-Gaussian~\cite{garcia2011analysis,kenny2010bayesian,cai2020deep,wang2020simulation}. In this case,
a linear Gaussian model such as PLDA is not sufficient.
Nevertheless, designing a more complex generative model is not always ideal~\cite{villalba2017tied,li2020neural},
as it may jeopardize the model's generalization, another important concern for open-set verification tasks.

The NL form provides a novel decoupling approach for the above dilemma. The key idea is that we never assume that
the three scoring components are tied to a single generative process. Instead, we shall assume two generative processes,
one for the global distribution, and the other for the local. Specifically, we hope all the inference related
to unseen classes is based on a \emph{global model} which should be simple, in order to ensure generalization,
and the inference related to individual classes is based on a \emph{local model} which could be complex, in order to ensure accuracy.

Among the three components in the NL form, the enrollment and normalization components are related to inference for unseen classes
and so should be based on the global model,
while the prediction component is for representing the within-class property and thus should be based on the local model.
Intuitively, this decoupling approach can be understood as a multi-level representation for the data: a high-level representation
that describes the holistic property of the entire data, and a low-level representation that describes the details of a particular class.

\subsection{Decoupled PLDA}

As the first study of the decoupling approach, we test a simple case in this paper: the global model is a linear Gaussian,
and the local model is another independent Gaussian. This model, which we call decoupled PLDA, is very similar to the vanilla PLDA,
with the only exception that the prediction component is never based on the within-class distribution learned by the global model, but a
new Gaussian. Formally, assume the global model as follows:

\begin{equation}
p_g(\pmb{\mu}) = \mathcal{N} (\pmb{\mu}; \pmb{0}, \textnormal{diag}(\pmb{\epsilon}))
\end{equation}

\begin{equation}
p_g(\pmb{x}| \pmb{\mu}) = \mathcal{N} (\pmb{x}; \pmb{\mu},  \mathbf{I}),
\end{equation}

\noindent and the local model supposes that after a linear transform $\mathbf{M}$,
the within-class distribution follows a Gaussian:

\begin{equation}
p_l(\mathbf{M}\pmb{x}|\pmb{\mu}) = \mathcal{N} (\mathbf{M}\pmb{x}; \pmb{\mu}, \mathbf{I}),
\end{equation}

\noindent where $\mathbf{M}$ is a transform matrix.
The two models are integrated by the NL formulation, shown as follows:

\begin{equation}
 \emph{NL} = \frac{\int p_l(\mathbf{M}\pmb{x}|\pmb{\mu}) p_g(\pmb{\mu}|\pmb{x}_1, ..., \pmb{x}_n) \rm{d} \pmb{\mu}}{ p_g(\pmb{x}) }.
\end{equation}

The decoupled PLDA can be trained following the maximum likelihood (ML) criterion. The training involves two steps:

\begin{itemize}
\item \textbf{Global model training}: The same as training vanilla PLDA, following the ML criterion.
More details can be found in the original paper~\cite{ioffe2006probabilistic,prince2007probabilistic}.
\item \textbf{Local model training}: Optimize the linear transform $\mathbf{M}$ by maximizing the likelihood function as follows:
\end{itemize}

\begin{small}
\begin{eqnarray}
\mathcal{L}(\mathbf{M}) &=& \prod_k^K \prod_{i=1}^{n_k} \int p_l(\mathbf{M}\pmb{x}^k_i|\pmb{\mu}) p_g(\pmb{\mu} | \pmb{x}^k_1, ..., \pmb{x}^k_{n_k}) \rm{d} \pmb{\mu} \nonumber \\
&=& \prod_k^K  \prod_{i=1}^{n_k} \int  \mathcal{N}(\mathbf{M}\pmb{x}^k_i;\pmb{\mu},   \mathbf{I}) p_g(\pmb{\mu} | \pmb{x}^k_1, ..., \pmb{x}^k_{n_k}) \rm{d} \pmb{\mu} \nonumber \\
&=& \prod_k^K  \prod_{i=1}^{n_k} \mathcal{N} (\mathbf{M}\pmb{x}^k_i;  \frac{n_k\pmb{\epsilon}}{n_k \pmb{\epsilon} + 1 } \bar{\pmb{x}}_k, (1 + \frac{\pmb{\epsilon} }{n_k \pmb{\epsilon} + 1}) \mathbf{I}), \nonumber \\
\end{eqnarray}
\end{small}

where $\bar{\pmb{x}}_k$ is the average of the samples in the $k$-th class.

\subsection{Discussion}
\label{sec:ana}

\subsubsection{Difference from PLDA}

One may argue that since the decoupled PLDA and vanilla PLDA are trained using the same data and follow the same model assumption (both $p_g$ and $p_l$ are Gaussians), 
they should be identical when optimized, i.e. $\mathbf{M}=\mathbf{I}$. This intuition however is not correct, due to two main reasons. Firstly, since $p_l$ and $p_g$ are independent,
the posterior derived from $p_g$, i.e., $p_g(\pmb{\mu}|\pmb{x}_1^k, ..., \pmb{x}_{n_k}^k)$, cannot be regarded as the posterior of $\pmb{\mu}$ in the local model.
Instead, it should be regarded as a \emph{prior} for the distribution of $\pmb{\mu}$ for each data sample. With this prior, the local model is trained by  improving the likelihood of the data on the local model. 
This is the key point of the decoupled model, where we assume two random processes, rather than a single one as in the vanilla PLDA.

Secondly, when training the local model, we treat all the samples are independent, no matter whether they belong to the same class or not. 
This is clear from the form of $\mathcal{L}(\mathbf{M})$ presented in (10), where the integration over $\pmb{\mu}$
is conducted for each training sample. Note that if it is not the case, the objective would be as follows:
\begin{equation}
\mathcal{L}(\mathbf{M}) = \prod_k^K \int \left[ \prod_{i=1}^{n_k}  p_l(\mathbf{M}\pmb{x}^k_i|\pmb{\mu}) \right] p_g(\pmb{\mu} | \pmb{x}^k_1, ..., \pmb{x}^k_{n_k}) \rm{d} \pmb{\mu}.
\end{equation}

The reason of assuming independent samples is that it matches the scenario in the \emph{test} phase, where all the test samples are independent. The objective for
the local model, therefore, essentially encourages positive samples obtaining higher likelihood with the hypothesized class means (inferred by the global model) 
during test.

\subsubsection{Single-set training}

One design choice for the decoupled PLDA is to split the training data into an enrollment set and a test set, and train the global and local models using the two sets
respectively. This would simulate the enroll-test scenario in a more realistic way. In this study, we choose a \emph{single-set} scheme where the global and local models
are trained using the same dataset, partly due to the limited training data in our experiments. 

\subsubsection{Early stop for unideal normalization}

A key problem of the decoupled PLDA is that with the increased likelihood $p_l(\mathbf{x}|\pmb{\mu})$ computed by the local model, the normalization term $p_g(\mathbf{x})$ computed by the global model
will be not theoretically correct. In this preliminary study, an empirical solution was used to address this problem: we monitor the performance
of the model on the \emph{training set} when training the local model, and stop the training process when the performance reaches the best. The resultant model trades off the
gain between the empowered local model and the loss caused by the unideal normalization. More theoretical solution will be left for future work.

\section{Related work}

Numerous techniques have been proposed to alleviate the deficiency of vanilla PLDA in representing complicated data.
For example, the heavy-tailed PLDA~\cite{kenny2010bayesian} uses a Student's $t$ distribution to handle long-tail distributions.
Mixture PLDA was proposed to handle the diverse statistics under different conditions~\cite{mak2015mixture,xie2020mixture,senoussaoui2011mixture}.
Deep normalization was proposed to gaussianize the distribution of speaker vectors~\cite{cai2020deep},
which was further extended to a neural discriminant analysis (NDA)~\cite{li2020neural}.
Although promising, few of the techniques can improve modeling capacity while maintaining simplicity.

\begin{table*}[htb]
  \caption{EER(\%) results with PLDA and dePLDA.}
  \label{tab:res}
  \centering
  \scalebox{1.0}{
  \begin{tabular}{l|lccc|lccc}
    \toprule
    \multicolumn{1}{c|}{} &    \multicolumn{4}{|c|}{i-vector}  &  \multicolumn{4}{|c}{x-vector} \\
    \midrule
     Model       &  Dim      &   SITW.Eval       &  AISHELL-1       &  HI-MIA               & Dim     &SITW.Eval       &  AISHELL-1       &  HI-MIA \\
    \midrule
      PLDA       &  400       &  8.201           &  1.179           &  1.008                & 512     & 6.862          &  1.146           &  0.698  \\
      PLDA       &  128       &  7.381           &  1.372           &  1.318                & 128     & 5.604          &  1.368           &  0.930   \\
    \midrule
      dePLDA     &  400       &  7.627           &  \textbf{1.075}  &  \textbf{0.698}       & 512     & 5.820          &  \textbf{0.792}  &  \textbf{0.465}\\
      dePLDA     &  128       &  \textbf{6.998}  &  1.278           &  1.240                & 128     & \textbf{4.950} &  1.028           &  0.775 \\
    \bottomrule
  \end{tabular}}
\end{table*}

\section{Experiments}

We evaluate decoupled PLDA by a speaker verification task. Speaker verification (SV) targets for determining
whether a speech segment is spoken by a claimed speaker or not~\cite{campbell1997speaker,reynolds2002overview,hansen2015speaker}.
Modern SV approaches firstly convert variable-length speech segments to fixed-dimensional vectors (a.k.a., speaker vectors), and then
employ PLDA to score the speaker vectors. 

Two types of speaker vectors are most popular: i-vector~\cite{dehak2011front} and x-vector~\cite{snyder2018xvector}. The former is
derived based on a statistical model, while the latter is based on deep neural networks~\cite{deng2014deep,ehsan14,li2017deep}.
In spite of other explorations presented recently~\cite{heigold2016end,zhang2016end,rahman2018attention,Cai2018,Jung2019raw,Wang2019phonetic,Wang2019centroid},
the i-vector/x-vector plus PLDA architecture is widely recognized as the state-of-the-art, partly due to the standard recipe published in the Kaldi toolkit~\cite{povey2011kaldi}.
In this paper, we will evaluate decoupled PLDA (dePLDA for short) with both i-vector and x-vector, and compare its performance with that of vanilla PLDA.

\subsection{Data}

\noindent \textbf{VoxCeleb}~\cite{nagrani2017voxceleb,chung2018voxceleb2}: An open-source speaker dataset collected from media sources by University of Oxford.
This dataset contains 2,000+ hours of speech signals from 7,000+ speakers.
This dataset was used to train the i-vector and x-vector models,
as well as the PLDA model used in the test on the SITW dataset.

\noindent \textbf{SITW}~\cite{mclaren2016speakers}: A standard evaluation dataset consists of 299 speakers.
The core-core trials built on the SITW.Dev and SITW.Eval sets were used for evaluation.


\noindent \textbf{AIShell-1}~\cite{aishell_2017}: An open-source Chinese Mandarin speech dataset.
The training set (used for PLDA and decoupled PLDA training) involves $360,897$ utterances from $340$ speakers, and the test set
involves $64,495$ utterances from $60$ speakers.

\noindent \textbf{HI-MIA}~\cite{qin2020hi}: An open-source text-dependent speaker recognition dataset.
All the speech utterances contain the word `Hi MIA', recorded by a microphone $1$ meter away from the speaker.
The training set (used for PLDA and decoupled PLDA training) involves $15,186$ utterances from $254$ speakers, and the test set involves $5,048$ utterances from $86$ speakers.

\subsection{Experimental setup}

The i-vector and x-vector models were implemented using the Kaldi toolkit~\cite{povey2011kaldi},
following the SITW recipe.
Although the recipe does not produce the best performance known so far on the SITW dataset, it is the most popular and reproducible, and so suitable for concept demonstration in our preliminary study.

The dimensionality of i-vectors and x-vectors was set to $400$ and $512$, respectively.
The PLDA and dePLDA models were trained based on the produced i-vectors and x-vectors.
PLDA was trained using the Kaldi toolkit, and decoupled PLDA was implemented by PyTorch\footnote{https://gitlab.com/csltstu/enroll-test-mismatch}.
The transform matrix $\mathbf{M}$ in decoupled PLDA was constrained to be diagonal and was initialized as an identity matrix.
The Adam optimizer~\cite{kingma2014adam} was used to optimize $\mathbf{M}$, and the best $\mathbf{M}$ is selected using
SITW.Dev for the SITW test, and their own training sets for the AIShell-1 test and HI-MIA test.
The selection is based on the performance in terms of the equal error rate (EER).

\subsection{Training process}

\begin{figure}[htb]
  \centering
  \includegraphics[width=0.80\linewidth]{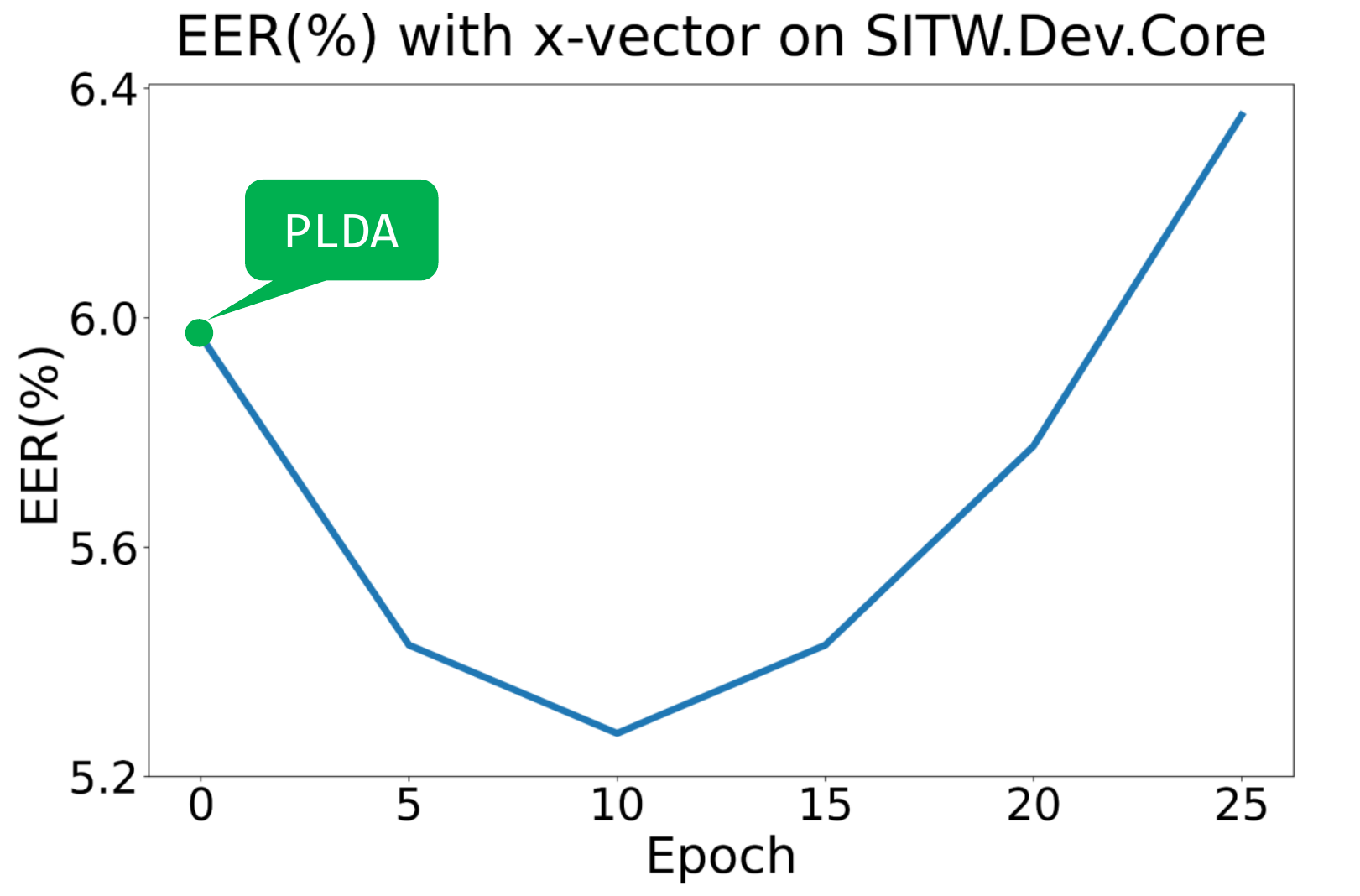}
  \caption{The trend of the EER results in the dePLDA training process. Note that at the beginning $\mathbf{M}$ is initialized as $\mathbf{I}$, which is equivalent to 
  the vanilla PLDA.}
  \label{fig:train}
\end{figure}

The change of the EER results on SITW.Dev core-core trials with dePLDA trained on VoxCeleb are reported in Fig.~\ref{fig:train}. 
It can be seen that the performance was improved with the training proceeding, and reached the best result after 10 iterations. After that, more iterations 
lead to performance degradation. This is the expected pattern, and reflects the trade-off between the empowered local model and the mismatch between 
the local and global models. 

\subsection{Main results}

The EER results on SITW.Eval, AISHELL-1 and HI-MIA are reported in Table~\ref{tab:res}. Two configurations were tested: 
(1) in the full-dimensional case, the PLDA/dePLDA models were trained with the raw speaker vectors; 
(2) in the low-dimensional case, the dimensionality of the speaker vectors was firstly reduced to 128 by LDA (as recommended in the Kaldi recipe), and then
the PLDA/dePLDA models were trained using the low-dimensional speaker vectors. 
The results show that with both the configurations, dePLDA
outperforms PLDA in a consistent way, demonstrating that the decoupling approach is rational and effective. Note that
the dimension reduction does not improve the performance in the AISHELL-1 and HI-MIA tests, which we attribute to the unreliable
LDA model with the limited training data, as well as the suboptimal dimensionality (128) that was tuned on SITW.

\subsection{Inconsistence with length normalization}

Length normalization (LN)~\cite{garcia2011analysis} is a standard preprocessing in speaker verification. It 
projects speaker vectors to a spherical surface, so that the (marginal) distribution tends to be more Gaussian. This
simple operation has been demonstrated highly effective for both i-vector systems and x-vector systems~\cite{garcia2011analysis}.
Unfortunately, LN and dePLDA are inconsistent,
as the local model shrinks the speaker vectors, which may counteract the effect of LN.

A possibility is to use a \emph{partial} LN, i.e., conducting LN within the global model only.
This is not theoretically correct as it may cause additional mismatch between the local model and the global model. 
Nevertheless, we found the partial LN may lead to performance gains in some cases.

\begin{table}[htb]
  \caption{EER(\%) results with PLDA and dePLDA based on the i-vector model.}
  \label{tab:deplda_iv}
  \centering
  \scalebox{1.0}{
  \begin{tabular}{lllccc}
    \toprule
    Model        &  Dim      &   LN        &  SITW.Eval              &  AISHELL-1       &  HI-MIA  \\   
    \midrule
     PLDA       &  400       &           &  8.201           &  1.179           &  1.008   \\
     PLDA       &  400       &   $\checkmark$        &  6.315           &  1.071           &  \textbf{0.698}   \\
     PLDA       &  128       &              &  7.381           &  1.372           &  1.318   \\
     PLDA       &  128       &  $\checkmark$       &  \textbf{5.878}  &  1.335           &  1.008   \\
    \midrule
     dePLDA     &  400       &          &  7.627           &  1.075           &  \textbf{0.698}   \\
     dePLDA     &  400       &   $\checkmark$      &  6.014           &  \textbf{1.056}  &  0.775   \\
     dePLDA     &  128       &        &  6.998           &  1.278           &  1.240   \\
     dePLDA     &  128       &  $\checkmark$       &  5.987           &  1.349           &  0.775   \\
    \bottomrule
  \end{tabular}}
\end{table}

\begin{table}[htb]
  \caption{EER(\%) results with PLDA and dePLDA based on the x-vector model.}
  \label{tab:deplda_xv}
  \centering
  \scalebox{1.0}{
  \begin{tabular}{lllccc}
    \toprule
    Model        &  Dim      &   LN        &  SITW.Eval              &  AISHELL-1       &  HI-MIA  \\
    \midrule
    PLDA       &  512       &           &  6.862            &  1.146           &  0.698   \\
    PLDA       &  512       &   $\checkmark$        &  4.511            &  0.806           &  0.698   \\
    PLDA       &  128       &              &  5.604            &  1.368           &  0.930   \\
    PLDA       &  128       &  $\checkmark$        &  \textbf{3.636}   &  1.019           &  0.620   \\
    \midrule
    dePLDA     &  512       &          &  5.820            &  0.792           &  \textbf{0.465}   \\
    dePLDA     &  512       &   $\checkmark$      &  4.292            &  \textbf{0.740}  &  0.853   \\
    dePLDA     &  128       &        &  4.950            &  1.028           &  0.775   \\
    dePLDA     &  128       &  $\checkmark$       & 4.019            &  1.198           &  0.620   \\
    \bottomrule
  \end{tabular}}

\end{table}

The results with i-vectors and x-vectors are shown in Table~\ref{tab:deplda_iv} and Table~\ref{tab:deplda_xv} respectively.
It can be seen that LN consistently improved the performance of the vanilla PLDA in all situations. The partial LN,
however, lead to improved performance with dePLDA in some cases, but in other cases, the performance was degraded. 
Accompanied by LN, vanilla PLDA generally outperformed dePLDA, although in some situations dePLDA accompanied by partial LN 
showed better performance (e.g., results on SITW with 512-dim x-vectors or 400-dim i-vectors).

The loss of the comparative advantage of dePLDA when LN is employed can be attributed to the improved Gaussianality of the length-normalized data.
Notice that if the data is truly linear Gaussian and perfectly matches the PLDA assumption, the vanilla PLDA is optimal. The decoupling
approach can contribute only if the data is complex and deviate from linear Gaussian.
LN arguably regulates the speaker vectors and makes them more Gaussian, which may reduce the potential contribution of the decoupling approach.

\section{Conclusions}

This paper reported a preliminary study towards decoupled scoring for open-set verification tasks, where a global model is designed to
provide a bird view for the entire profile of the data, and a local model to provide a detailed view for individual classes.
To verify its effectiveness, we tested a simple decoupled PLDA model (dePLDA),
where both the global and local models are Gaussian. Compared to the vanilla PLDA, the new model uses a separate Gaussian to
match the within-class distribution, and simulates sample independence in the test phase. 

We verify the decoupled PLDA model with a speaker verification task on three datasets. The results show that working on raw speaker vectors, 
the decoupled PLDA
consistently outperformed the vanilla PLDA, confirming that the decoupling approach is effective. However, when length normalization is
employed, the comparative advantage of dePLDA was largely lost. This is probably attributed to the improved Gaussianality of the speaker vectors 
after length normalization, which diminishes the potential of the decoupling approach.
Nevertheless, our experiments indeed demonstrated that the decoupling approach is rational and effective, and the performance might be significantly improved 
if the local model is more expressive, especially when it is non-Gaussian.

\section*{Acknowledgment}

This work was supported by the National Natural Science Foundation of China (NSFC) under Grants No.61633013 and No.62171250, 
and also the Huawei Innovation Research Program under Cross-Device Speaker Recognition Project Contract No.YBN2019125091.

\bibliographystyle{IEEEtran}
\bibliography{refs}

\end{document}